\documentclass[12pt,thmsa]{article}
\usepackage{sw20lart}

\input tcilatex
\begin{document}

\author{I. Radinschi\thanks{%
iradinsc@phys.tuiasi.ro} \and Department of Physics, ``Gh. Asachi''
Technical University, \and Iasi, 6600, Romania}
\title{On the Bergmann Energy-Momentum Complex of a Charged Regular Black Hole}
\date{March 29, 2001}
\maketitle

\begin{abstract}
We use the Bergmann energy-momentum complex to calculate the energy of a
charged regular black hole. The energy distribution is the same as we
obtained in the Einstein prescription. Also, we get the expression of the
energy in the Bergmann prescription for a general spherically symmetric
space-time of the Kerr-Schild class.

Keywords: energy momentum-complex, charged regular black hole

PACS: 04.20.Dw, 04.70.Bw
\end{abstract}

\section{Introduction}

The subject of energy-momentum localization in general relativity continues
to be an open one because there is no given yet a generally accepted
expression for the energy-momentum density. Even they are coordinate
dependent various energy-momentum complexes give the same energy
distribution for a given space-time. Aguirregabiria, Chamorro and Virbhadra 
\cite{1} obtained that the energy-momentum complexes of Einstein \cite{2},
Landau and Lifshitz \cite{3}, Papapetrou \cite{4} and Weinberg \cite{5} give
the same result for the energy distribution for any Kerr-Schild metric.
Also, recently Virbhadra \cite{6} investigated if these definitions lead to
the same result for the most general non-static spherically symmetric metric
and found they don't give the same expression for the energy distribution.
He concluded that only the energy-momentum complex of Einstein still gives
the same energy distribution when the calculations are performed in the
Kerr-Schild Cartesian and Schwarzschild Cartesian coordinates.

Chang, Nester and Chen \cite{7} showed that the energy-momentum complexes
are actually quasilocal and legitimate expression for the energy-momentum.
Very important is the Cooperstock \cite{8} hypothesis which states that the
energy and momentum are confined to the regions of non-vanishing
energy-momentum tensor of the matter and all non-gravitational fields.

In this paper we calculate the energy distribution of a charged regular
black hole using the Bergmann energy-momentum complex \cite{9}. Also, we
obtain the expression of the energy in the Bergmann prescription for a
general static spherically symmetric space-time of the Kerr-Schild class. We
show that the Bergmann energy-momentum complex is a good tool for evaluating
the energy distribution. We use geometrized units $\left( G=1\text{, }%
c=1\right) $ and follow the convention that Latin indices run from 0 to 3.

\section{Energy in the Bergmann prescription}

E. Ay\'{o}n-Beato and A. Garcia (ABG) \cite{10} gave recently a solution to
the coupled system of the Einstein field and equations of the nonlinear
electrodynamics. This is a singularity free black hole solution with mass $M$
and electric charge $q$. Also, the metric at large distance behaves as the
Reissner-Nordstr\"{o}m (RN) solution. The usual singularity of the RN
solution, at $r=0$, has been smoothed out and now it simply corresponds to
the origin of the spherical coordinates. The line element is given by 
\begin{equation}
ds^2=A\left( r\right) dt^2-B\left( r\right) dr^2-r^2\left( d\theta ^2+\sin
^2\theta d\varphi ^2\right) ,  \tag{1}
\end{equation}
where 
\begin{equation}
A\left( r\right) =B^{-1}\left( r\right) =1-\frac{2M}r\left( 1-\tanh \left( 
\frac{q^2}{2Mr}\right) \right) .  \tag{2}
\end{equation}

If the electric charge vanishes we reach the Schwarzschild solution. At
large distances (2) resembles to the Reissner-Nordstr\"{o}m solution and can
be written 
\begin{equation}
A\left( r\right) =B^{-1}(r)=1-\frac{2M}r+\frac{q^2}{r^2}-\frac{q^6}{12M^2r^4}%
+O\left( \frac 1{r^6}\right) .  \tag{3}
\end{equation}

We show that the solution given by (1) can be transformed to another form of
a general spherically symmetric space-time of the Kerr-Schild class. For the
Kerr-Schild class space-times the metrics $g_{ik}$ have the form

\begin{equation}
g_{ik}=\eta _{ik}-Hl_il_k,  \tag{4}  \label{4}
\end{equation}

where $\eta _{ik}=diag\left( 1,-1,-1,-1\right) $ is the Minkowski metric. $H$
represents the scalar field and $l_i$ is a null, geodesic and shear free
vector field in the Minkowski space-time. We also have 
\begin{equation}
\begin{array}{l}
\eta ^{ab}l_al_b=0, \\ 
\eta ^{ab}l_{i,a}l_b=0, \\ 
\left( l_{a,b}+l_{b,a}\right) l_{\;\,,c}^a\eta ^{bc}-\left(
l_{\;\,,a}^a\right) ^2=0.
\end{array}
\tag{5}  \label{5}
\end{equation}

For the metric given by (1) we make the transformation 
\begin{equation}
u=t+\int A^{-1}\left( r\right) dr  \tag{6}  \label{6}
\end{equation}
and we have 
\begin{equation}
dt=du-A^{-1}\left( r\right) dr.  \tag{7}  \label{7}
\end{equation}

We obtain 
\begin{equation}
dt^2=du^2+A^{-2}\left( r\right) dr^2-2A^{-1}\left( r\right) drdu.  \tag{8}
\label{8}
\end{equation}

With (8) the metric given by (1) becomes 
\begin{equation}
ds^2=A\left( r\right) du^2-2dudr-r^2\left( d\theta ^2+\sin ^2\theta d\varphi
^2\right) .  \tag{9}  \label{9}
\end{equation}

This metric corresponds to the particular static case of the general
non-static spherically symmetric space-time of the Kerr-Schild class
considered by Virbhadra \cite{6} (see Eq. (28) in his paper), for the
evaluation of the energy in the Einstein, Landau and Lifshitz, Papapetrou
and Weinberg prescriptions.

For the line element (9) we can use the transformations 
\begin{equation}
\begin{array}{l}
u=T+r, \\ 
x=r\sin \theta \cos \varphi , \\ 
y=r\sin \theta \sin \varphi , \\ 
z=r\cos \theta
\end{array}
\tag{10}
\end{equation}
and we obtain 
\begin{equation}
ds^2=dT^2-dx^2-dy^2-dz^2-\left( 1-A\right) \times \left[ dT+\frac{xdx+ydy+zdz%
}r\right] ^2.  \tag{11}
\end{equation}

This is a Kerr-Schild class metric with $H=1-A$ and $l_i=\left( 1,\dfrac
xr,\dfrac yr,\dfrac zr\right) $.

The Bergmann energy-momentum complex \cite{9} is given by 
\begin{equation}
B^{ik}=\frac 1{16\pi }\frac{\partial \left( g^{il}H_l^{\;km}\right) }{%
\partial x^m}  \tag{12}
\end{equation}

where 
\begin{equation}
H_l^{\;km}=\frac{g_{\ln }}{\sqrt{-g}}\left[ -g\left(
g^{kn}g^{mp}-g^{mn}g^{kp}\right) \right] _{,p}  \tag{13}
\end{equation}

and with $H^{ikm}=g^{il}H_l^{\;km}$. The energy and momentum are given by 
\begin{equation}
P^i=\iiint B^{i0}dx^1dx^2dx^3.  \tag{14}
\end{equation}

Using the Gauss theorem we have 
\begin{equation}
P^i=\frac 1{16\pi }\iint H^{i0\alpha }\,n_\alpha \,dS,  \tag{15}
\end{equation}
where $n_\alpha =\left( \dfrac xr,\dfrac yr,\dfrac zr\right) $ are the
components of a normal vector over an infinitesimal surface element $%
dS=r^2\sin \theta d\theta d\varphi $.

For the metric given by (11) we obtain the nonvanishing $H^{00\alpha }$
components 
\begin{equation}
\begin{array}{l}
H^{001}=\dfrac{-2x\left( -1+A\left( r\right) \right) }{r^2}, \\ 
\\ 
H^{002}=\dfrac{-2y\left( -1+A\left( r\right) \right) }{r^2}, \\ 
\\ 
H^{003}=\dfrac{-2z\left( -1+A\left( r\right) \right) }{r^2}.
\end{array}
\tag{16}
\end{equation}

Now, using (15) and (16) we get the energy distribution for the space-time
described by (9) 
\begin{equation}
E\left( r\right) =\frac r2\left( 1-A\left( r\right) \right) .  \tag{17}
\end{equation}

For the (ABG) black hole, with (2) and (17) we obtain 
\begin{equation}
E\left( r\right) =M\left( 1-\tanh \left( \frac{q^2}{2Mr}\right) \right) 
\tag{18}
\end{equation}
and 
\begin{equation}
E\left( r\right) =M-\frac{q^2}{2r}+\frac{q^6}{24r^3M^2}-\frac{q^{10}}{%
240M^4r^5}+O\left( \frac 1{r^6}\right) .  \tag{19}
\end{equation}

Also, (19) can be written 
\begin{equation}
E\left( r\right) =E_{RN}\left( r\right) +\frac{q^6}{24r^3M^2}-\frac{q^{10}}{%
240M^4r^5}+O\left( \frac 1{r^6}\right) .  \tag{20}
\end{equation}

The term $E_{RN}\left( r\right) $ represents the energy of the
Reissner-Nordstr\"{o}m solution that corresponds to the Penrose \cite{25}
quasi-local mass definition.

With the notations $E^{^{\prime }}=\frac{E(r)}M$, $Q=\frac qM$ and $R=\frac
rM$ we have $E^{^{\prime }}=1-\tanh (\frac{Q^2}{2R})$. We plot the
expression of $E^{^{\prime }}$ in FIGURE\ 1 ($E^{^{\prime }}$ on Z-axis is
plotted against $R$ on X-axis and $Q$ on Y-axis).

\section{Discussion}

Bondi \cite{24} gave his opinion that a nonlocalizable form of energy is not
admissible in relativity.

Many results recently obtained \cite{11}-\cite{23} sustain that the
energy-momentum complexes can give reasonable results.

Chang, Nester and Chen \cite{7} argued that every energy-momentum complex is
associated with a legitimate Hamiltonian boundary term, and, because of this
the energy-momentum complexes are quasilocal and acceptable. Each is the
energy-momentum density for some physical situation. This Hamiltonian
approach to quasilocal energy-momentum rehabilitates the energy-momentum
complexes.

Using an adequate coordinate transformation we get for the line element (1)
the form given by (9). For this form of metric we calculate the energy
distribution with the energy-momentum complex of Bergmann. We use
Kerr-Schild Cartesian coordinates. The expression of the energy is the
particular static case of the result obtained by Virbhadra \cite{6} (see Eq.
(31) therein), for a general non-static spherically symmetric space-time of
the Kerr-Schild class, using the Einstein, Landau and Lifshitz, Papapetrou
and Weinberg prescriptions. Also, in his earlier paper Virbhadra \cite{1}
showed that several energy-momentum complexes give the same result for a
Kerr-Schild class metric. Tod obtained the same expression for the energy as
obtained by Virbhadra using the Penrose \cite{25} quasi-local mass
definition (see in \cite{6}).

For the (ABG) charged regular black hole the energy distribution calculated
in the Bergmann prescription depends on the mass $M$ and electric charge $q$%
. The first two terms in the expression of the energy represent the energy
distribution of the Reissner-Nordstr\"{o}m solution. The other terms are due
to the nonlinearity effect. The result is the same as we get in the Einstein
prescription \cite{26} for the metric given by (1) and (2).

We made the calculations using Kerr-Schild Cartesian coordinates and the
Bergmann energy-momentum complex provides for the metric given by (9) the
same expression for the energy distribution as the Einstein, Landau and
Lifshitz, Papapetrou and Weinberg energy-momentum complexes. Our result
sustain the importance of the energy-momentum complexes in the evaluation of
the energy distribution of a given space-time.

\textbf{Acknowledgments}

I am grateful to Professor K. S. Virbhadra for his helpful advice.

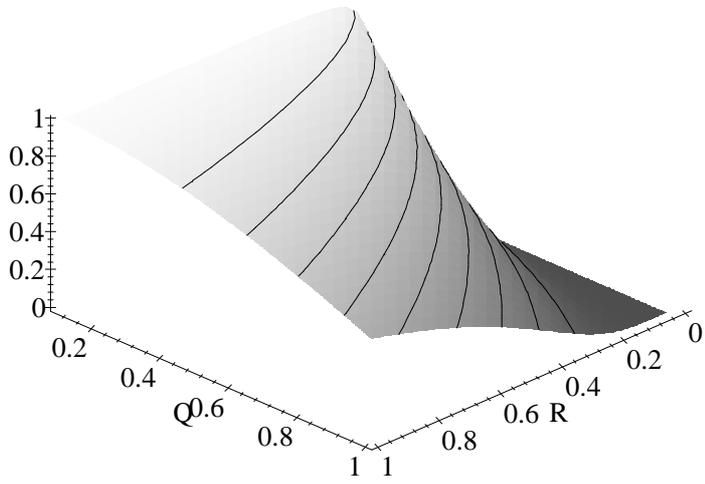
\end{document}